\documentclass[conference]{IEEEtran}
\IEEEoverridecommandlockouts
\usepackage[T1]{fontenc}

\newif\ifblind
\blindfalse   % \blindtrue for anonymized submission; \blindfalse for camera-ready / internal circulation

\usepackage{cite}
\usepackage{amsmath,amssymb,amsfonts}
\usepackage{algorithmic}
\usepackage{graphicx}
\usepackage{textcomp}
\usepackage{xcolor}
\usepackage{booktabs}
\usepackage{tabularx} % flexible-width tables that stay within a column
\usepackage{url}
\usepackage[hidelinks]{hyperref}
\usepackage{fancyhdr}   % camera-ready copyright footer (no TikZ overlay)
\usepackage{enumitem} % allows itemize/enumerate optional args like [leftmargin=*, itemsep=..., topsep=...]

\graphicspath{{figures/}}

\fancypagestyle{sc26copyright}{%
  \fancyhf{}%
  \fancyfoot[L]{\fontsize{7}{8}\selectfont
    \shortstack[l]{%
      SC26 Workshops, November 15--20, 2026, Chicago, Illinois, USA\\
      979-8-3195-1221-5/26/\$31.00 \textcopyright{}2026 IEEE
    }% 
  }%
}

\begin{document}

\title{From Pilots to Production: Lessons in\\
Cross-Institutional Federated Training and Artificial Intelligence for Science}

% ----- Authors -----------------------------------------------------------
\ifblind
  \author{\IEEEauthorblockN{Anonymous Author(s)}
  \IEEEauthorblockA{Affiliation withheld for double-blind review}}
\else
  \author{
  \IEEEauthorblockN{Olivera Kotevska\IEEEauthorrefmark{1},
    Max Carlson\IEEEauthorrefmark{2},
    Yan Gao\IEEEauthorrefmark{4},
    Francis Jeanson\IEEEauthorrefmark{6},
    Yijiang Li\IEEEauthorrefmark{3},
    %William Lindskog\IEEEauthorrefmark{5},\\
    %Mohammad Manzari\IEEEauthorrefmark{6},
    Mohammad Naseri\IEEEauthorrefmark{4},\\
    %Holger Roth\IEEEauthorrefmark{4},
    Minseok Ryu\IEEEauthorrefmark{5},
    Sahil Tyagi\IEEEauthorrefmark{1},
    Jerry Watkins\IEEEauthorrefmark{2},
    Feiyi Wang\IEEEauthorrefmark{1},
    Ravi Madduri\IEEEauthorrefmark{3},
    Kibaek Kim\IEEEauthorrefmark{3}}
  \IEEEauthorblockA{
    \IEEEauthorrefmark{1}Oak Ridge National Laboratory, \quad
    \IEEEauthorrefmark{2}Sandia National Laboratories, \quad
    \IEEEauthorrefmark{3}Argonne National Laboratory,\\
    %\IEEEauthorrefmark{4}NVIDIA, \quad
    \IEEEauthorrefmark{4}Flower Labs, \quad
    %\IEEEauthorrefmark{6}Deloitte, \quad
    \IEEEauthorrefmark{5}Arizona State University, \quad
    \IEEEauthorrefmark{6}Ontario Brain Institute
  }
}
\fi

\maketitle
% ----- Required IEEE copyright footer (camera-ready only) ----------------
\ifblind\else
\thispagestyle{sc26copyright}
\fi

% ----- Abstract ----------------------------------------------------------
\begin{abstract}
Many of the most valuable scientific datasets cannot be centralized: they are proprietary, export-controlled, classified, or bound by data-sovereignty restrictions. This inverts the usual paradigm: the model must move to the data, making federated artificial intelligence (AI) core infrastructure for open science. We synthesize lessons from U.S. Department of Energy national laboratories, industry deployments, and the open-source community into a structured account of moving federated AI from pilot demonstrations to dependable, multi-site production. The lessons come from five concurrent efforts spanning leadership-class supercomputers and cloud infrastructure in regulated settings. We organize them around an adapted \textit{five-domain readiness frame} and a stratified view of the stack beneath a trained model: data architecture, privacy and security controls, trust and verification, governance and socio-technical factors, and operations, the layers that decide whether a pilot becomes dependable. The question shifts from ``can we train it?'' to ``can we operate it, audit it, and change it safely?'' We report systems lessons in memory efficiency, reliability, and synchronization; set out what privacy, security, and decentralized trust require in production, and what is not yet validated there; and identify cross-cutting open problems (asynchronous federation, leakage auditing, verification standards, and harmonized data contracts) that we argue warrant a dedicated, international, open-science working group.
\end{abstract}

\begin{IEEEkeywords}
federated learning, large language models, scientific AI, high-performance computing, privacy, trust, cross-institutional collaboration
\end{IEEEkeywords}

% ----- Body --------------------------------------------------------------
\section{Introduction}
\label{sec:intro}

Artificial Intelligence (AI) for science has shifted from building ever-larger models from scratch towards adapting pre-trained foundation models and composing them into systems that couple reasoning, domain expertise, agentic workflows, and experimental instruments. That shift puts data access at the center, yet some of the most valuable data is precisely what cannot move.

\subsection{The centralization barrier}
Across national laboratories, hospitals, pharmaceutical companies, and manufacturing facilities, vast datasets are proprietary, export-controlled, classified, or subject to the Health Insurance Portability and Accountability Act (HIPAA)~\cite{hipaa1996}, the General Data Protection Regulation (GDPR)~\cite{gdpr2016}, or institutional data-governance policies. Scale compounds the problem. Leadership-class high-performance computing (HPC) facilities hold the compute to train frontier AI models: Frontier, Aurora, Polaris, Perlmutter, and El Capitan, operated by the U.S. Department of Energy (DOE) computing facilities~\cite{olcf,alcf,nersc}. The data those models need is spread across institutions whose policies prevent centralized aggregation. The asymmetry holds outside the laboratory system: commercial cloud supplies elastic capacity and open-source frameworks orchestrate across both regimes, yet neither dissolves the policy boundaries that keep data in place. The response is to share the model, not the data~\cite{mcmahan2017communication}: each institution contributes the signal in its local data without the raw data crossing a security boundary.

\subsection{Federated AI across sectors and domains}
Federated learning (FL) makes this possible in principle, with a growing body of work across healthcare, finance, and scientific computing (Section~\ref{sec:related}). The lessons we report were developed in a national-laboratory setting, but the barriers they address arise in any cross-institutional federated deployment: memory pressure, reliability failures, synchronization bottlenecks, governance gaps, and trust deficits. The synthesis is deliberately cross-sector. Industry contributes cloud-based federations in regulated settings, which separate lessons specific to leadership-class facilities from those that hold anywhere (Section~\ref{sec:beyondhpc}), and the socio-technical account of why federations stall on contracting and incentives rather than compute (Section~\ref{sec:frame}). The open-source community contributes the frameworks themselves, whose developers co-developed the aggregation-memory fixes with laboratory teams, and the decentralized verification model of Section~\ref{sec:privacy}. Together these let the lessons travel beyond the DOE.

\subsection{From pilots to production}
A successful pilot is not a production system: a trained model is one artifact in a multi-site deployment that must also be governed, secured, operated, and recovered across institutions sharing neither hardware nor trust assumptions~\cite{roth2026production}. Moving from ``can we train it?'' to ``can we operate it, audit it, and change it safely?'' is the central challenge, and at frontier scale it demands systems that sustain the memory, reliability, and synchronization requirements of models with tens to hundreds of billions of parameters across heterogeneous facilities.

\subsection{Contributions}

\begin{table}[t]
\caption{The five efforts synthesized here span four sectors and two infrastructure regimes. Sections~\ref{sec:systems} and~\ref{sec:privacy} draw on all five.}
\label{tab:efforts}
\scriptsize
\setlength{\tabcolsep}{3pt}
\begin{tabularx}{\columnwidth}{@{}lXX@{}}
\toprule
\textbf{Effort} & \textbf{Sector / infrastructure} & \textbf{Primary contribution} \\
\midrule
NNSA tri-lab & DOE laboratory; on-premise leadership-class HPC & Memory pressure; cross-vendor portability \\
ASCR cross-facility & DOE laboratory; co-scheduled leadership-class HPC & Synchronization; queue-aware aggregation \\
NeuroFL & Non-profit health research; cloud & Domain extension under regulated data \\
Commercial deployments & Consulting; cloud & Socio-technical readiness \\
Privacy optimization & Academia; controlled evaluation & Privacy and utility tradeoff \\
\bottomrule
\end{tabularx}
\end{table}

This paper is a synthesis and characterization rather than a single-method paper. It draws on five concurrent, independent efforts, listed in Table~\ref{tab:efforts}, each resolving or sharpening a distinct bottleneck. Specifically, we:
\begin{itemize}
    \item Frame cross-institutional federated AI readiness across five domains, adapted from Roth et al.~\cite{roth2026production}, and map them onto a stratified production stack (Section~\ref{sec:frame});
    \item Report systems lessons on memory, reliability, and synchronization from two frontier-scale deployments: federated large language model (LLM) training across three National Nuclear Security Administration (NNSA) laboratories, and cross-facility fine-tuning across four DOE supercomputers (Section~\ref{sec:systems});
    \item Distill production best practices for privacy, security, and decentralized trust, and situate active privacy-preserving optimization against what is validated on deployed systems (Section~\ref{sec:privacy});
    \item Identify cross-cutting open problems and propose a dedicated Trillion Parameter Consortium (TPC) working group (Section~\ref{sec:open}).
\end{itemize}

The paper moves from frame to evidence to agenda: the readiness domains and the stack they are assessed against (Section~\ref{sec:frame}); what two frontier-scale deployments taught us and how far those lessons reach (Section~\ref{sec:systems}); what privacy, security, and trust require in production, and where the evidence runs out (Section~\ref{sec:privacy}); and what remains unsolved (Section~\ref{sec:open}).

\section{Related Work}
\label{sec:related}

\paragraph{Federated LLM training}
FL was formalized by McMahan et al.~\cite{mcmahan2017communication}; Kairouz et al.~\cite{kairouz2021advances} survey the cross-silo regime most relevant here. Federating LLM training compounds its challenges: weight sizes create memory pressure at aggregation, inter-facility bandwidth constrains communication, and long runs amplify reliability risk~\cite{ye2024openfedllm}. Federated LLM training across the NNSA laboratories is reported in~\cite{owens2025three} and was presented at NVIDIA FLARE Day~\cite{carlson2025nvflare}.

\paragraph{FL in industry and regulated sectors}
Cross-silo FL is furthest along outside science. Healthcare consortia train without pooling patient data~\cite{rieke2020future}, mature deployments exist in finance and manufacturing, and the pattern now appears in neuroscience networks~\cite{neurofl2026} and commercial deployments where contracting velocity, not compute, sets the pace~\cite{manzari2026sociotechnical}. These run on cloud infrastructure under health-privacy statute rather than facility policy, a useful control on lessons drawn from leadership-class systems (Section~\ref{sec:beyondhpc}).

\paragraph{FL on leadership-class high-performance computing (HPC)}
Li et al.~\cite{li-kim2025scalable,li-kim2026fedqueue} demonstrate cross-facility FL across four DOE supercomputers, fine-tuning a 7B model on 1{,}700+ graphics processing units (GPUs) from three vendors. Kotevska et al.~\cite{kotevska2026scalable} validate scalable FL for scientific foundation models on leadership-class systems, and survey the privacy challenges this raises for science~\cite{kim-kotevska2025privacy,kotevska2025privacy}. This paper synthesizes the systems lessons emerging across these concurrent efforts.

\paragraph{Privacy in FL}
Gradient-inversion and membership-inference attacks show that shared updates leak training data without raw data access~\cite{geiping2020inverting, zhu2019deep}. Differential privacy (DP)~\cite{dwork2006differential} gives the standard guarantee at a documented utility cost, whether applied per record~\cite{abadi2016deep} or per client~\cite{geyer2017differentially}; adaptive estimators~\cite{byeon-ryu2026dpstein} improve the tradeoff but are not yet validated at frontier scale (Section~\ref{sec:privacy}).

\paragraph{FL frameworks/ecosystems}
NVFlare~\cite{roth2022nvidia} provides a production-oriented platform; APPFL~\cite{ryu-kim2022appfl, li-kim2024advances} targets cross-facility HPC via Globus integration; Flower~\cite{beutel2020flower} emphasizes ecosystem breadth, with domain-specific extensions such as NeuroFL for neuroscience research~\cite{neurofl2026}; and OmniFed~\cite{tyagi-kotevska2025omnifed} decouples topology, communication backend, and privacy mechanism through configuration, addressing the heterogeneity that recurs throughout this paper. We use this landscape to motivate the readiness frame in Section~\ref{sec:frame}.

\section{A Common Frame: The Production Stack and Five Readiness Domains}
\label{sec:frame}

\begin{figure}[t]
    \centering
    \includegraphics[width=0.86\linewidth]{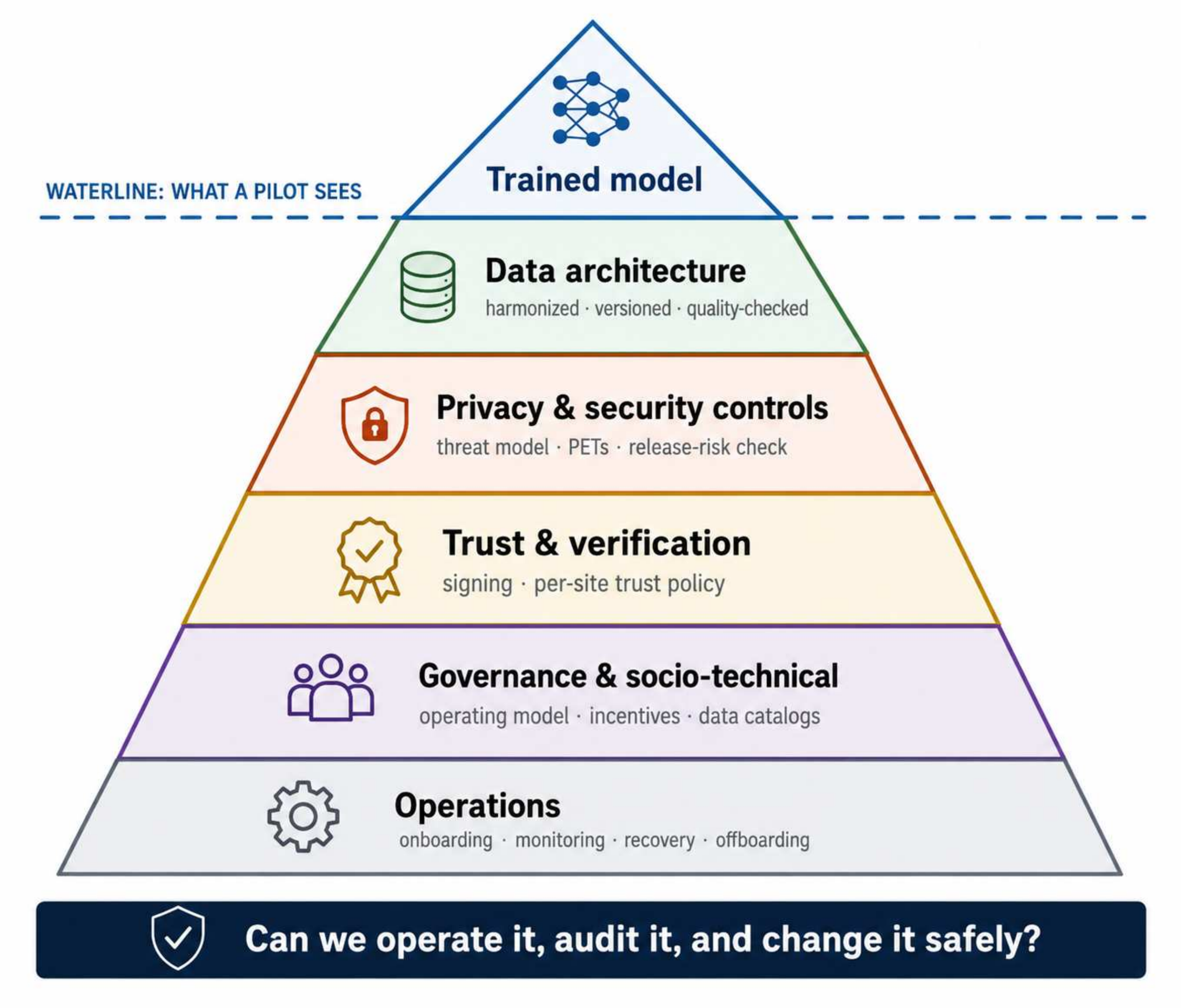}
    \caption{The production iceberg: the trained model above the waterline, and the five non-training strata beneath it, each labelled with the concerns it must cover.}
    \label{fig:iceberg}
\end{figure}

Disparate federated efforts are easier to compare against a common structure. Figure~\ref{fig:iceberg} gives one: the trained model is the visible tip a pilot sees, dependable production rests on a larger hidden mass, and each stratum below the waterline is a place a federation can fail.

\emph{Data architecture} keeps raw data local while making contributions comparable: a common data model with versioned extract-transform-load (ETL), quality control, and modality checks. The same schema does not guarantee the same meaning. \emph{Privacy and security controls} are chosen against an explicit threat model rather than adopted by default, with a release-risk check before anything leaves a site (Section~\ref{sec:privacy}). \emph{Trust and verification} asks how a site decides that code it did not write is safe to run on its own data; the answer is signing and a per-site trust policy (Section~\ref{sec:privacy}). \emph{Governance and socio-technical factors} are the operating model, incentives, and catalogs deciding what is actually shared. \emph{Operations} is onboarding, monitoring, recovery, and offboarding across sites that share neither hardware nor trust (Section~\ref{sec:systems}). Together they are the difference between a model that trains and a system that can be run.

The stack says what must exist. It does not say whether a given federation is ready to operate it. For that we adapt the five-domain readiness frame of Roth et al.~\cite{roth2026production}, renaming its domains so they cannot be mistaken for the strata above. The frame separates a pilot mindset (few sites, manual approvals, custom scripts) from a production mindset built on reusable onboarding, evidence-based approvals, versioned pipelines, and observable operations. Its discipline is to assess each domain with evidence, not opinion; a missing gap becomes a remediation plan or a reason not to scale yet. The domains are:
\begin{enumerate}
    \item \textbf{Purpose \& Value.} A measurable reason to exist: narrow testable questions, confirmed site-level data availability, pre-agreed success thresholds.
    \item \textbf{Agreements \& Approvals.} A durable operating agreement: accountable operating model, approval path, collaboration terms. Frequently the critical path in regulated domains.
    \item \textbf{Data Fitness.} Harmonized, validated, versioned inputs.
    \item \textbf{Risk \& Justification.} Controls chosen against explicit threats rather than adopted by default (Section~\ref{sec:privacy}).
    \item \textbf{Run \& Recover.} Reproducible, observable, resilient across sites, with approved runtimes from on-premise to HPC (Section~\ref{sec:systems}).
\end{enumerate}

Each domain answers for a part of the stack. The domains are what a federation \emph{assesses} before scaling; the strata are what it \emph{builds and operates}:

\begin{center}\small
\setlength{\tabcolsep}{2pt}
\begin{tabularx}{\columnwidth}{@{}r@{\ $\rightarrow$\ }>{\raggedright\arraybackslash}X@{}}
Purpose \& Value & no stratum; assessed before anything is built \\
Agreements \& Approvals & Governance \& socio-technical \\
Data Fitness & Data architecture \\
Risk \& Justification & Privacy \& security controls; Trust \& verification \\
Run \& Recover & Operations \\
\end{tabularx}
\end{center}

\noindent Risk \& Justification spans two strata because deciding which code is safe to run is a build concern distinct from protecting what leaves a site.

A purely technical assessment also misses the \emph{hidden iceberg} of socio-technical barriers~\cite{manzari2026sociotechnical}: federations most often fail for want of coordination across people, process, and technology. Forming one requires mapping institutional process velocity (contracting, procurement, legal approval, platform deployment) as carefully as the technology stack; practitioners need playbooks and sandboxes, not application programming interfaces (APIs) alone; and incentive misalignment between principal investigators and data custodians, where sharing follows social capital rather than policy, governs what data is available. Sections~\ref{sec:systems} and~\ref{sec:privacy} populate this frame with concrete results. The governing principle throughout is to \emph{use federation to reduce data movement, then use controls to manage the residual risk}~\cite{roth2026production}.

\section{Systems Lessons from Frontier-Scale Federated Training}
\label{sec:systems}

We ground the readiness frame in two independent frontier-scale deployments sharing one architecture: a central aggregation server coordinating clients on each site's login nodes, so model weights cross the network while raw data stays local across different DOE supercomputers, vendors, and workflow stacks. Both confirm that cross-facility federated LLM training is feasible at scale, and surface barriers that appear only under production conditions.

\subsection{Deployment I: tri-laboratory training on NNSA systems}
This deployment~\cite{carlson2026nnsa,owens2025three,carlson2025nvflare} federates Sandia National Laboratories (SNL), LLNL, and Los Alamos National Laboratory (LANL), spanning both major accelerator vendors: LLNL's El Capitan (AMD MI300A, No.~2 on the June 2026 TOP500 list at 1.809~Exaflop/s~\cite{top5002026june}), and SNL and LANL machines with NVIDIA H100s. Clients execute on login nodes, submit training as Slurm or Flux batch jobs, and communicate over client-initiated Secure Shell (SSH) tunnels; access-restricted data never leaves its originating site. The backend training is done primarily using torchtitan; a portable, HPC-oriented LLM stack supporting sharded data, tensor, pipeline, and context parallelism on both vendors. As part of this work, three federated training frameworks (NVFlare, Flower, and APPFL) were evaluated by doing continued pre-training of Llama 3.x models (1B, 8B, and 70B) on arXiv documents that were published after the Llama 3 training cutoff date and split evenly among the three sites.

\begin{figure}[t]
    \centering
    \includegraphics[width=0.88\linewidth]{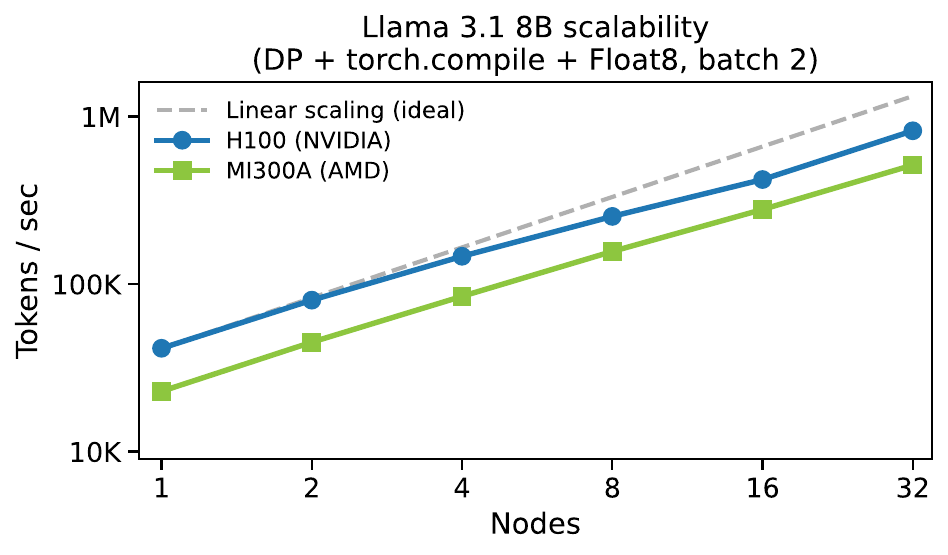}
    \caption{Llama 3.1 8B throughput vs.\ node count on NVIDIA H100 and AMD MI300A hardware, using a shared portable training backend (DP + torch.compile + Float8, batch 2).}
    \label{fig:scaling}
\end{figure}

Continued pre-training using torchtitan scaled well on both vendor architectures for the unclassified arXiv experiments. Figure~\ref{fig:scaling} shows per-node token throughput for the 8B model up to 32 nodes: both hardware types track close to linear scaling, with the AMD MI300A matching or slightly exceeding H100 throughput at larger node counts. All experiments have torch's compile option enabled and 8-bit floating-point (Float8) precision.

This workflow has been made production-ready and deployed on larger classified machines including El Capitan, and subsequently used to train models on proprietary data spread across the three NNSA labs.

\subsection{Deployment II: cross-facility training on ASCR systems}
\begin{figure}[t]
    \centering
    \includegraphics[width=0.88\linewidth]{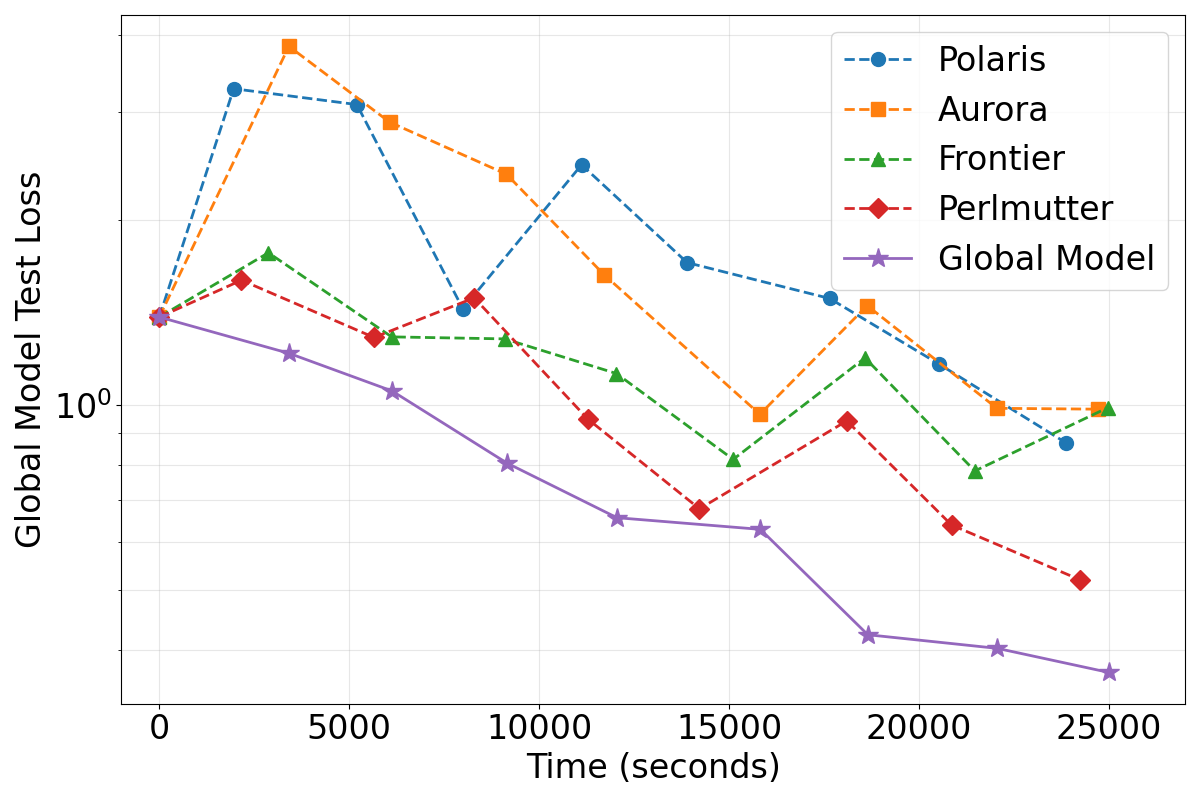}
    \caption{Per-facility and global-model test loss for cross-facility fine-tuning of LLaMA-2 7B (SMolInstruct, chemistry-subdomain non-IID split) on Aurora, Polaris, Frontier, and Perlmutter via APPFL and Globus.}
    \label{fig:testloss}
\end{figure}

An independent deployment on Office of Advanced Scientific Computing Research (ASCR)~\cite{ascr} systems using APPFL and Globus infrastructure~\cite{li-kim2025scalable} demonstrates cross-facility federated fine-tuning of a 7B model across four DOE ASCR supercomputers (Aurora and Polaris at Argonne, Frontier at Oak Ridge, Perlmutter at Berkeley) spanning three GPU vendors (Intel, NVIDIA, and AMD). Using a co-scheduled 64-node-per-facility reservation (1{,}700+ GPUs total), LLaMA-2 7B is fine-tuned on SMolInstruct, a chemistry instruction-tuning dataset partitioned by task group across facilities to induce a natural non-independent and identically distributed (non-IID) split that mirrors real institutional specialization (e.g., reaction chemistry at Aurora, property prediction at Polaris). Global test loss drops monotonically from 1.39 to 0.37 over eight aggregation rounds (Figure~\ref{fig:testloss}), consistently outperforming every individual facility's local model, evidence that the federated model integrates complementary domain knowledge without centralizing raw data. A companion small-scale study without HPC reservations, where jobs instead enter each facility's real production queue, further shows that batch-queue delays vary substantially and unpredictably across facilities and can dominate wall-clock training time. These observations motivate FedQueue~\cite{li-kim2026fedqueue}, a queue-aware protocol that treats scheduler admission delay as a first-class signal and reaches a 20\% better final loss and 60\% faster time to target accuracy than FedAvg, FedAsync, FedBuff, and FedCompass (Figure~\ref{fig:fedqueue}, Artifact Description appendix).

\subsection{Three systems lessons}
Three lessons recurred across both deployments and the framework evaluations around them. \textit{First}, memory pressure at aggregation was the first bottleneck: a 70B model requires ${\sim}130$~GB per copy in 32-bit precision, straining shared login nodes. It is now tractable through message quantization, native tensor transfer, and file-based streaming developed jointly by laboratory teams and framework developers. \textit{Second}, reliability must be designed in: as models grow and runs lengthen, framework-level persistent state becomes a first-class requirement so runs can be resumed rather than restarted from scratch. \textit{Third}, synchronization is the next frontier: existing frameworks' synchronous server/client communication dominates failure and wasted wall-clock time at scale, motivating the asynchronous or relaxed-synchronization protocols we return to in Section~\ref{sec:open}.

\subsection{How far the lessons generalize}
\label{sec:beyondhpc}
Both deployments share one regime: on-premise leadership-class systems, batch-scheduled clients, co-located trust anchors. Whether the lessons are facts about federation or about that regime, the laboratory deployments alone cannot say. Two efforts in Table~\ref{tab:efforts} operate under the opposite constraints~\cite{neurofl2026,manzari2026sociotechnical}. Both are cloud-hosted rather than HPC-based, communicate over encrypted client-server channels, and handle health data governed by statute rather than by facility security policy: HIPAA~\cite{hipaa1996}, Ontario's Personal Health Information Protection Act (PHIPA)~\cite{phipa2004}, or Canada's Personal Information Protection and Electronic Documents Act (PIPEDA)~\cite{pipeda2000}.

Statute, not site policy, binds here, and the change of regime separates facility artifacts from durable ones. Memory pressure follows from shared login nodes and dissolves when the aggregator is an elastic cloud instance. Reliability and synchronization persist in altered form: cloud clients do not queue, but they churn, and round completion is still gated by the least-available participant. The socio-technical findings of Section~\ref{sec:frame} were most pronounced in these regulated settings. Our base nonetheless remains concentrated in DOE laboratories, with no deployment at a National Science Foundation (NSF) center~\cite{nsf,accessci}; these share an allocation process, a security model, and often a software stack, so some lessons may reflect that substrate rather than federation itself. Extending to NSF and EuroHPC~\cite{eurohpc} facilities is a priority we return to in Section~\ref{sec:open}. With the systems constraints in view, we next turn to privacy, security, and trust controls for production deployments.

\section{Privacy, Security, and Trust: What Production Requires}
\label{sec:privacy}

Federation reduces data movement but not risk: updates, gradients, metrics, and outputs can all leak sensitive information even when raw data never leaves a site~\cite{kim-kotevska2025privacy}. Production-grade federated AI therefore has to treat privacy, security, and trust as layered controls mapped to explicit threats and release risks~\cite{roth2026production}. A companion perspective develops this in depth, treating privacy as an assurance problem whose protected unit is often an institution rather than a record~\cite{kotevska2026foundations}.

\subsection{Defense in depth}
A useful organizing structure asks three questions of any federated deployment~\cite{roth2026production}:
\begin{itemize}
    \item \textbf{Who can connect?} Identity, authorization, certificate and key management determine which participants may join the federation at all.
    \item \textbf{What can move?} Updates, metrics, and outputs need privacy-enhancing technologies (PETs) matched to the threat model; a systematic FL threat taxonomy~\cite{shi-kotevska2024dealing} is the right starting point. The PET toolbox layers encryption, secure aggregation, differential privacy, and confidential computing so no single mechanism is a single point of failure~\cite{carlson2026nnsa}.
    \item \textbf{Where can it run?} Trusted execution and attestation give higher-assurance guarantees for the most sensitive environments.
\end{itemize}

\subsection{Make controls an evidence-based decision}
Layered controls are credible only if their selection is justified. A discipline that recurred across efforts is a three-part evidence chain: a threat model naming the attack scenarios and sensitivities at stake; a control rationale justifying which PETs and runtime controls that threat model warrants; and a release-risk check asking, before anything is shared, whether it is safe to release the model, its metrics, or its outputs~\cite{roth2026production}. Controls should match the threats rather than be applied indiscriminately. In the tri-laboratory deployment only model weights cross inter-site connections, and the residual risk of sharing them is what the surrounding controls and release reviews manage.

\subsection{Decentralized trust for shared apps}
Organizations increasingly run others' code on their own sensitive data, and each client with access to the global model poses a leakage risk~\cite{xu-kotevska2026tramark}. How does a site know what runs on its client, and decide what to trust~\cite{lindskog2026flower}? Two common answers scale poorly: per-site self-review duplicates effort and yields inconsistent security review, while a single centralized reviewer creates both a bottleneck and a single point of trust. Decentralized verification offers a better tradeoff: publishers and reviewers sign applications, a hub distributes them with their signatures, and each site applies its own policy over reviewers it trusts~\cite{lindskog2026flower}. Paired with a stable, versioned application format, discoverability, reusability, and ``safe to run'' become properties of the ecosystem rather than burdens on each participant.

\subsection{Where the evidence runs out}
A persistent tension underlies all of the above: privacy controls such as differential privacy can materially degrade model utility, and the strength of the guarantee is often coupled to the loss in accuracy. Active work seeks to improve this tradeoff through better statistical estimation under privacy noise~\cite{byeon-ryu2026dpstein}, gradient subspace learning~\cite{kotevska2026dptwolevel}, and heterogeneous client-level budgets~\cite{xu-kotevska2026gdpfed}. We are careful not to overclaim: to date these are demonstrated in controlled settings rather than on deployed frontier-scale systems, often at privacy budgets that should be reported plainly.

\section{What Remains Unsolved, and a Shared Agenda}
\label{sec:open}

The problems that remain are best tackled collectively.

\begin{itemize}[leftmargin=*,itemsep=2pt,topsep=3pt]
    \item \textbf{Asynchronous and relaxed-synchronization federation.} Asynchronous protocols that preserve convergence and privacy accounting while tolerating stragglers and site failures are, in our view, the most consequential open systems problem here.
    \item \textbf{Leakage auditing on deployed systems.} Best practices presume a threat model, but quantifying residual leakage from shared weights under realistic multi-site conditions~\cite{shi-kotevska2024dealing} is unvalidated at frontier scale, as is tracing leakage to a responsible participant~\cite{xu-kotevska2026tramark}.
    \item \textbf{Standardized verification and trust.} Decentralized verification (Section~\ref{sec:privacy}) points toward a scalable trust model, but the community lacks shared standards for application formats, signatures, attestation, and trust-policy expression, standardization that would reduce duplicated security review.
    \item \textbf{Harmonized data contracts and governance.} Federations need versioned, validated data contracts (common data models, quality and drift signals, compatibility checks) that travel with the federation rather than being re-negotiated per project. The organizational infrastructure of Section~\ref{sec:frame} is a precondition.
    \item \textbf{Privacy-aware optimization at scale.} How adaptive aggregation, heterogeneous budgets, and communication-efficient methods interact at cross-facility scale remains open.
    \item \textbf{Foundation-model scaling and portability.} At trillion-parameter scale, full-weight exchange becomes impractical, requiring low-rank adapters, sparse deltas, and revised privacy accounting; sustaining one portable backend across three vendors is a parallel cost.
\end{itemize}

\smallskip\noindent\textbf{Toward a working group.}\label{sec:workinggroup} National laboratories, industry, and the open-source community arrived at these problems independently, the clearest evidence that none belongs to one institution. A dedicated Trillion Parameter Consortium working group~\cite{tpc2026workshop} could coordinate the work, set shared evidence standards, and keep the results open.

\clearpage
\section*{Acknowledgments}
This material is based upon work co-supported by the U.S.\ Department of Energy, Office of Science, Office of Advanced Scientific Computing Research under Contract No.\ DE-AC05-00OR22725. This manuscript has been co-authored by UT-Battelle, LLC under Contract No.\ DE-AC05-00OR22725 with the U.S.\ Department of Energy. The United States Government retains and the publisher, by accepting the article for publication, acknowledges that the United States Government retains a non-exclusive, paid-up, irrevocable, world-wide license to publish or reproduce the published form of this manuscript, or allow others to do so, for United States Government purposes. The Department of Energy will provide public access to these results of federally sponsored research in accordance with the DOE Public Access Plan (\url{http://energy.gov/downloads/doe-public-access-plan}). \par
This work was co-supported by the U.S.\ Department of Energy, Office of Science, Advanced Scientific Computing Research, under Contract DE-AC02-06CH11357. An award of computer time was provided by the ASCR Leadership Computing Challenge (ALCC) program. This research used resources of the Argonne Leadership Computing Facility, which is a U.S.\ Department of Energy Office of Science User Facility operated under contract DE-AC02-06CH11357. This research used resources of the Oak Ridge Leadership Computing Facility at the Oak Ridge National Laboratory, which is supported by the Office of Science of the U.S.\ Department of Energy under Contract No.\ DE-AC05-00OR22725. This research used resources of the National Energy Research Scientific Computing Center (NERSC), a Department of Energy User Facility using NERSC award ALCC-ERCAP0038201. We gratefully acknowledge the computing resources provided on Improv, a high-performance computing cluster operated by the Laboratory Computing Resource Center at Argonne National Laboratory. \par
For the NNSA lab federated training portion of this work, the authors would like to thank Nick Winovich (SNL) and Sarah Tsai (SNL) for help with dataset curation. Additionally, the authors would like to thank Yang Ho (SNL), Nicholas Sly (LLNL), Josh Kallman (LLNL), and Scott Pakin (LANL) for their contributions towards developing the tri-lab model and Charles Jekel for help running torchtitan on AMD hardware. Finally, the authors would like to thank Chris Siefert (SNL), Sivasankaran Rajamanickam (SNL), and Daniel O'Malley (LANL) for overall project guidance.
\par

% ----- Bibliography ------------------------------------------------------
\bibliographystyle{IEEEtran}
\bibliography{references}

% ----- Appendices (do NOT count toward the 5-page limit) -----------------
%\appendices
\section*{Artifact Description (AD) Appendix}
\label{sec:appendix}

This appendix follows the SC26 AD Appendix structure. Because this paper is a synthesis of five concurrent efforts rather than a single-method paper, the artifacts described here are the deployment and evaluation artifacts underlying the two frontier-scale case studies of Section~\ref{sec:systems} and the demonstrator implementations behind the privacy and trust discussion of Section~\ref{sec:privacy}. Presentation materials, curated code pointers, and companion documentation are collected at the project repository, \url{https://github.com/ORNL/RealWorld_FL}.

\subsection*{Part 1: Overview of Contributions and Artifacts}

\subsubsection*{A. Paper's Main Contributions}
\begin{enumerate}[label=C\arabic*.,leftmargin=*]
    \item A five-domain readiness frame for cross-institutional federated AI, adapted from Roth et al.~\cite{roth2026production}, and the stratified production stack it is assessed against (Section~\ref{sec:frame}).
    \item Systems lessons from two independent frontier-scale federated LLM deployments, covering memory efficiency, reliability, and synchronization (Section~\ref{sec:systems}).
    \item What privacy, security, and decentralized trust require in production, spanning national-laboratory deployments and industry framework practice, and what is not yet validated there (Section~\ref{sec:privacy}).
    \item Identification of what remains unsolved, framed as a shared international, open-science agenda (Section~\ref{sec:open}).
\end{enumerate}

\subsubsection*{B. Computational Artifacts}
Contributions C2 and C3 are supported by the computational artifacts listed in Table~\ref{tab:ad-artifacts}. Contributions C1 and C4 are conceptual and do not rely on computational reproduction.

\begin{table}[t]
\centering
\caption{Summary of computational artifacts supporting Contributions C2 and C3.}
\label{tab:ad-artifacts}
\footnotesize
\setlength{\tabcolsep}{3pt}
\renewcommand{\arraystretch}{1.1}
\begin{tabularx}{\linewidth}{@{}p{0.10\linewidth}X p{0.22\linewidth}p{0.14\linewidth}@{}}
\hline
\textbf{Artifact} & \textbf{Description} & \textbf{Reproduces} & \textbf{Contrib.} \\
\hline
A1 & NNSA tri-laboratory federated LLM training (NVFlare + \emph{torchtitan}) & Figure~\ref{fig:scaling} & C2 \\
A2 & Cross-facility federated fine-tuning (APPFL + Globus) across four ASCR supercomputers & Figure~\ref{fig:testloss} & C2 \\
A3 & FedQueue queue-aware FL protocol & Figure~\ref{fig:fedqueue} & C2 \\
A4 & TraMark demonstrator for traceable model-leakage attribution & Section~\ref{sec:privacy} discussion & C3 \\
A5 & GDPFed demonstrator for heterogeneous client-level differential privacy & Section~\ref{sec:privacy} discussion & C3 \\
\hline
\end{tabularx}
\end{table}

\subsection*{Part 2: Artifact Identification}

\subsubsection*{Artifact A1: NNSA Tri-Laboratory Federated LLM Training}

\textbf{A. Relation to contributions.} A1 grounds C2 with an operational cross-vendor federated training system spanning three NNSA laboratories, demonstrating throughput scaling and cross-vendor portability.

\textbf{B. Expected results.} Reproducing A1 yields per-node token-throughput measurements for Llama 3.x continued pre-training on H100 and MI300A hardware. Both vendors track close to linear scaling from 1 to 32 nodes (Fig.~\ref{fig:scaling}).

\textbf{C. Expected reproduction time (minutes).} Setup: 60. Execution: 240 (unclassified subset). Analysis: 15.

\textbf{D. Artifact setup.}
\begin{itemize}
    \item \emph{Hardware.} NVIDIA H100 GPUs (SNL, LANL) and AMD MI300A accelerated processing units (APUs) at LLNL's El Capitan. Reproducibility on the unclassified subset requires equivalent H100- or MI300A-class hardware.
    \item \emph{Software.} NVFlare (v2.5+), APPFL, Flower; \emph{torchtitan} (backend trainer); PyTorch with \emph{torch.compile}; Float8 support. Slurm or Flux scheduler; client-initiated SSH tunneling.
    \item \emph{Datasets.} arXiv documents published after the Llama 3 training cutoff (available via arXiv bulk access), split evenly across three sites.
    \item \emph{Installation.} See project repository for environment definitions and site-specific launcher scripts. Reproduction of the classified portion is not available outside NNSA facilities.
\end{itemize}

\textbf{E. Artifact evaluation.} Workflow: (T1) prepare split arXiv corpus $\rightarrow$ (T2) launch federated training round on each site via NVFlare $\rightarrow$ (T3) aggregate weights across sites over SSH tunnel $\rightarrow$ (T4) log per-node token throughput. Parameters: 8B Llama 3.1 model, batch size 2, Float8 precision, node counts $\{1,2,4,8,16,32\}$, eight aggregation rounds.

\textbf{F. Artifact analysis.} Raw throughput logs are aggregated per node count and plotted on log-log axes against the ideal linear-scaling reference to produce Fig.~\ref{fig:scaling}.

\subsubsection*{Artifact A2: Cross-Facility Federated Fine-Tuning on ASCR Systems}

\textbf{A. Relation to contributions.} A2 grounds C2 with a co-scheduled cross-facility fine-tuning deployment across four DOE ASCR supercomputers and three GPU vendors.

\textbf{B. Expected results.} Global model test loss decreases monotonically from approximately 1.39 to 0.37 over eight aggregation rounds, consistently outperforming every individual facility's local model (Fig.~\ref{fig:testloss}).

\textbf{C. Expected reproduction time (minutes).} Setup: 90. Execution: 420 (full co-scheduled run). Analysis: 20.

\textbf{D. Artifact setup.}
\begin{itemize}[leftmargin=*]
    \item \emph{Hardware.} 64 nodes each on Aurora (Intel Ponte Vecchio), Polaris (NVIDIA A100), Frontier (AMD MI250X), and Perlmutter (NVIDIA A100). Total 1{,}700+ GPUs under co-scheduled reservation.
    \item \emph{Software.} APPFL (v1.2+), Globus Compute, Globus Transfer, PyTorch with vendor-specific accelerator backends, LLaMA-2 7B base weights.
    \item \emph{Datasets.} SMolInstruct chemistry instruction-tuning dataset (publicly available), partitioned by task group across the four facilities to induce a natural non-IID split.
    \item \emph{Installation.} See APPFL documentation~\cite{li-kim2024advances} and companion scripts in the project repository.
\end{itemize}

\textbf{E. Artifact evaluation.} Workflow: (T1) partition SMolInstruct by task group $\rightarrow$ (T2) initialize APPFL server and per-facility Globus Compute endpoints $\rightarrow$ (T3) execute eight aggregation rounds with LLaMA-2 7B fine-tuning $\rightarrow$ (T4) collect per-facility and global test-loss trajectories. Parameters: LLaMA-2 7B, non-IID partition by chemistry subdomain, 8 aggregation rounds, standard FedAvg aggregation.

\textbf{F. Artifact analysis.} Test-loss trajectories from each facility and the global model are plotted against wall-clock time to produce Fig.~\ref{fig:testloss}. Full co-scheduled reproduction requires Innovative and Novel Computational Impact on Theory and Experiment (INCITE) or ASCR Leadership Computing Challenge (ALCC) allocations at all four facilities; a smaller-scale variant without HPC reservations is available for reviewers.

\subsubsection*{Artifact A3: FedQueue Queue-Aware FL Protocol}

\textbf{A. Relation to contributions.} A3 grounds C2 by isolating the impact of scheduler-admission latency and demonstrating that queue-aware aggregation improves both accuracy and time-to-target.

\begin{figure}[t]
    \centering
    \includegraphics[width=0.88\linewidth]{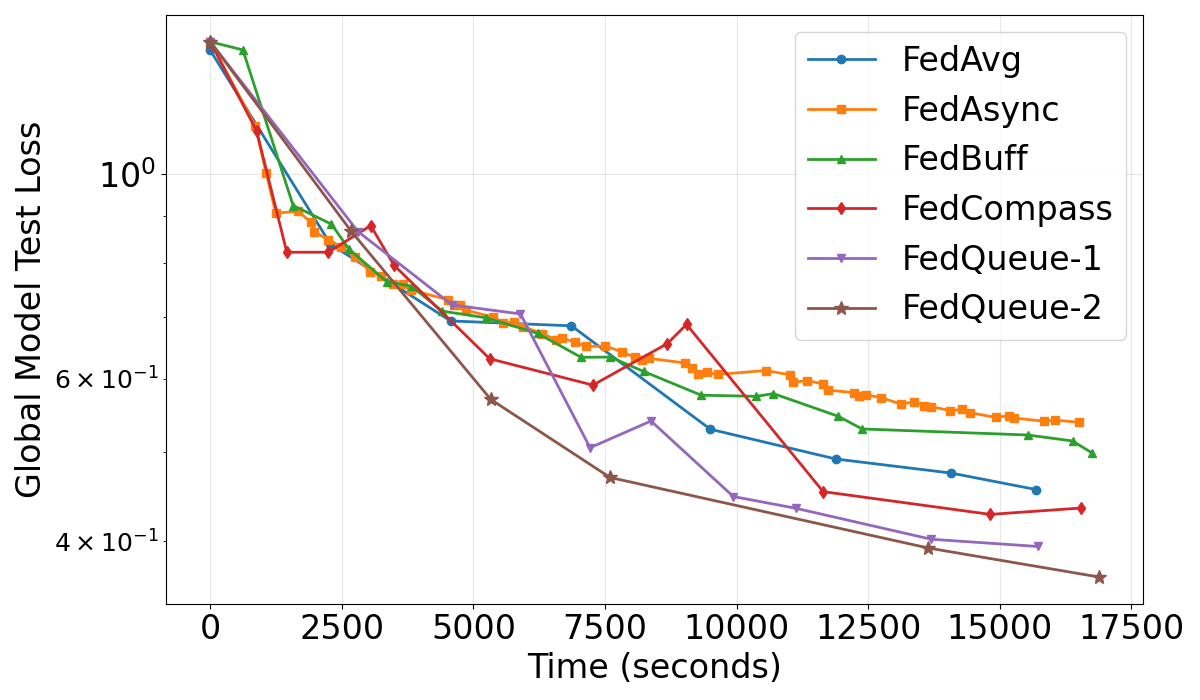}
    \caption{Global test loss against wall-clock time under heterogeneous batch-queue delays, comparing scheduler-aware aggregation (FedQueue) with FedAvg, FedAsync, FedBuff, and FedCompass.}
    \label{fig:fedqueue}
\end{figure}

\textbf{B. Expected results.} FedQueue reaches lower global test loss in wall-clock time than FedAvg, FedAsync, FedBuff, and FedCompass under heterogeneous batch-queue delays, yielding a 20\% better final loss and 60\% faster time to target accuracy (Fig.~\ref{fig:fedqueue}).

\textbf{C. Expected reproduction time (minutes).} Setup: 20. Execution: 60. Analysis: 10.

\textbf{D. Artifact setup.}
\begin{itemize}[leftmargin=*]
    \item \emph{Hardware.} Single node with GPU sufficient for controlled queue simulation; no HPC reservation required.
    \item \emph{Software.} APPFL with FedQueue extension; baseline strategies (FedAvg, FedAsync, FedBuff, FedCompass).
    \item \emph{Datasets.} Same SMolInstruct partition used in A2.
    \item \emph{Installation.} Reference implementation and simulation harness released with~\cite{li-kim2026fedqueue}.
\end{itemize}

\textbf{E. Artifact evaluation.} Workflow: (T1) load empirical queue-delay distributions from ASCR facility production traces $\rightarrow$ (T2) execute controlled simulation over each FL strategy $\rightarrow$ (T3) log global loss vs.\ wall-clock time. Parameters: five FL strategies, matched aggregation budget, three repetitions per strategy.

\textbf{F. Artifact analysis.} Loss-vs.-time trajectories are averaged over repetitions and plotted to produce Fig.~\ref{fig:fedqueue}. Time-to-target-accuracy is computed from smoothed curves.

\subsubsection*{Artifact A4: TraMark Demonstrator}

\textbf{A. Relation to contributions.} A4 substantiates C3 by providing a reference implementation of traceable black-box watermarking for client-level model leakage attribution, one of the two active-research directions surfaced in Section~\ref{sec:privacy}.

\textbf{B. Expected results.} TraMark achieves near-perfect leakage-tracing verification without material main-task accuracy degradation~\cite{xu-kotevska2026tramark}.

\textbf{C. Expected reproduction time (minutes).} Setup: 10. Execution: 45 (CIFAR-10, 10 clients). Analysis: 5.

\textbf{D. Artifact setup.}
\begin{itemize}[leftmargin=*]
    \item \emph{Hardware.} Single GPU workstation (any recent NVIDIA GPU with $\geq$16~GB memory).
    \item \emph{Software.} PyTorch; reference implementation at \url{https://github.com/JiiahaoXU/TraMark}.
    \item \emph{Datasets.} CIFAR-10 (publicly available; downloaded automatically via standard torchvision loaders).
    \item \emph{Installation.} Standard \texttt{pip install} of dependencies listed in the repository.
\end{itemize}

\textbf{E. Artifact evaluation.} Workflow: (T1) prepare CIFAR-10 with $N$ clients under a non-IID Dirichlet partition $\rightarrow$ (T2) run federated training with TraMark server-side watermark injection $\rightarrow$ (T3) query watermarked models via a black-box application programming interface (API) and compute verification rate. Parameters: 10 clients, $\alpha=0.04$, $k=0.01$, non-IID with $\gamma=0.5$.

\textbf{F. Artifact analysis.} Verification rate and main-task accuracy tabulated across configurations and compared to the baselines described in~\cite{xu-kotevska2026tramark}.

\subsubsection*{Artifact A5: GDPFed Demonstrator}

\textbf{A. Relation to contributions.} A5 substantiates C3 by providing a reference implementation of heterogeneous client-level differential privacy with optimized client sampling, the second of the two active-research directions surfaced in Section~\ref{sec:privacy}.

\textbf{B. Expected results.} GDPFed and GDPFed$^+$ yield higher utility than DP-FedAvg under heterogeneous per-client privacy budgets on Fashion MNIST (FMNIST) and CIFAR-10 benchmarks~\cite{xu-kotevska2026gdpfed}.

\textbf{C. Expected reproduction time (minutes).} Setup: 10. Execution: 60 (FMNIST and CIFAR-10). Analysis: 10.

\textbf{D. Artifact setup.}
\begin{itemize}[leftmargin=*]
    \item \emph{Hardware.} Single GPU workstation (any recent NVIDIA GPU with $\geq$16~GB memory).
    \item \emph{Software.} PyTorch; reference implementation at \url{https://github.com/JiiahaoXU/GDPFed}.
    \item \emph{Datasets.} FMNIST, CIFAR-10, and Shakespeare (all publicly available).
    \item \emph{Installation.} Standard \texttt{pip install} of dependencies listed in the repository.
\end{itemize}

\textbf{E. Artifact evaluation.} Workflow: (T1) partition the dataset into three privacy-budget groups with configurable per-group $\varepsilon$ $\rightarrow$ (T2) run P-FedAvg, DP-FedAvg, GDPFed, and GDPFed$^+$ over matched aggregation rounds using the sampling ratios computed from the paper's optimization formulation $\rightarrow$ (T3) log test accuracy and effective per-group $\varepsilon$ consumption. Reference privacy budgets in the paper are $\varepsilon_2=1.5, \varepsilon_3=3.0$ for FMNIST and $\varepsilon_2=6.0, \varepsilon_3=12.0$ for CIFAR-10, with $\varepsilon_1$ varied to examine sampling-ratio dynamics.

\textbf{F. Artifact analysis.} Convergence curves and test accuracy tabulated against baselines; per-group sampling ratios plotted against $\varepsilon_1$.

\bigskip

\noindent\emph{Reproduction access.} Reproduction scope for each artifact, including which parts are reproducible without facility access, is given in the Artifact Evaluation appendix.

\section*{Artifact Evaluation (AE) Appendix}
\label{sec:ae_appendix}

Because this paper is a synthesis of five concurrent efforts rather than a report of a single set of new experiments, the results shown here (Figures~\ref{fig:scaling}, \ref{fig:testloss}, \ref{fig:fedqueue}, and the privacy demonstrators of Section~\ref{sec:privacy}) are drawn from constituent works whose experimental protocols and validation studies are documented in their primary publications. The AD Appendix gives the setup, execution, and analysis path for each artifact. This appendix does not repeat them. It records provenance: for each artifact, where the reported result originates, where the code and data live, and what a reviewer can independently reproduce. Companion pointers and curated documentation for all five artifacts are collected at \url{https://github.com/ORNL/RealWorld_FL}.

\subsection*{Provenance and reproduction scope}
For each artifact below, \emph{primary source} is where the reported result was originally produced and validated; this paper reproduces or summarizes it.

\begin{itemize}[leftmargin=*,itemsep=3pt,topsep=3pt]
\item \textbf{A1 (Fig.~\ref{fig:scaling}).} Primary source: Carlson et al.~\cite{carlson2026nnsa}; Carlson and Siefert~\cite{carlson2025nvflare}; Owens~\cite{owens2025three}. Code and data: project repository; arXiv bulk access. Reproduction: unclassified arXiv subset only, on H100- or MI300A-class hardware. Production runs on El Capitan and the classified NNSA machines cannot be reproduced outside those facilities.
\item \textbf{A2 (Fig.~\ref{fig:testloss}).} Primary source: Li et al.~\cite{li-kim2025scalable}. Code and data: APPFL releases~\cite{li-kim2024advances}; SMolInstruct (public). Reproduction: the full run requires co-scheduled INCITE or ALCC allocations at four facilities simultaneously; a small-scale variant using each facility's production queue is public.
\item \textbf{A3 (Fig.~\ref{fig:fedqueue}).} Primary source: Li et al.~\cite{li-kim2026fedqueue}. Code and data: APPFL fork released with the same work. Reproduction: full, by controlled queue simulation on a single node; no HPC reservation needed.
\item \textbf{A4 (Section~\ref{sec:privacy}).} Primary source: Xu et al.~\cite{xu-kotevska2026tramark}. Code and data: \url{https://github.com/JiiahaoXU/TraMark}; CIFAR-10. Reproduction: full, on a single GPU workstation.
\item \textbf{A5 (Section~\ref{sec:privacy}).} Primary source: Xu et al.~\cite{xu-kotevska2026gdpfed}. Code and data: \url{https://github.com/JiiahaoXU/GDPFed}; FMNIST, CIFAR-10, Shakespeare. Reproduction: full, on a single GPU workstation.
\end{itemize}

\subsection*{What this paper contributes to each result}
For A1, A2, and A3 the figures are reproduced from the reference deployments; the numerical evaluation, ablations, and analytic bounds behind them are reported in the primary sources listed above. This paper adds the cross-effort synthesis, not new measurements. For A4 and A5 no figure is reproduced: the demonstrators are cited in Section~\ref{sec:privacy} as reviewable implementations of the two active research directions discussed there, and their validation is likewise documented in the primary sources.

\subsection*{Limits on independent verification}
Two artifacts cannot be fully reproduced by a reviewer. A1 depends on NNSA site access for the classified portion, and the unclassified arXiv subset requires H100- or MI300A-class hardware. A2 depends on co-scheduled leadership-class allocations at four facilities simultaneously; the qualitative trend is reproducible at small scale without reservations. A3, A4, and A5 require only a single GPU and are reproducible in full. We state this explicitly rather than implying uniform reproducibility across the artifact set.

\end{document}